\documentclass[fleqn,10pt]{wlscirep}
\usepackage[utf8]{inputenc}
\usepackage[T1]{fontenc}
\usepackage{amsmath}
\usepackage{fancyhdr}
\usepackage{scrextend}
\title{Concepts of Architecture, Structure and System\footnote{This material is part of both a research paper and a post graduate textbook in process of publication. Permission is granted for use by ISO for standards development. Other use requires written permission.}}
\author[1,2,$\dagger$]{Prof Michael K. Wilkinson}
\author[2,$\dagger$$\dagger$]{Prof Charles E. Dickerson}
\author[2,3,$\dagger$$\dagger$$\dagger$]{Dr Siyuan Ji}
\affil[1]{Atkins, Royal Pavilion, Wellesley Road, Aldershot, Hampshire, UK}
\affil[2]{School of Mechanical, Electrical and Manufacturing Engineering, Loughborough University, Loughborough, UK}
\affil[3]{Department of Computer Science, University of York, York, UK}
\affil[$\dagger$]{mike.wilkinson@atkinsglobal.com}
\affil[$\dagger\dagger$]{c.dickerson@lboro.ac.uk}
\affil[$\dagger\dagger\dagger$]{siyuan.ji@york.ac.uk}
\begin{abstract}
The current ISO standards pertaining to the Concepts of System and Architecture express succinct definitions of these two key terms that lend themselves to practical application and can be understood through elementary mathematical foundations. The current work of the ISO/IEC Working Group 42 is seeking to refine and elaborate the existing standards. This position paper revisits the conceptualisation underlying both of these key terms and offers an approach to: (i) refine and exemplify the term ‘fundamental concepts’ in the current ISO definition of Architecture; (ii) exploit existing standards and mathematical formalisms for the term ‘concept’; (iii) introduce a mathematical basis for the concept of Structure that can serve to unify the current terminology; and (iv) demonstrate that from a mathematical viewpoint the current definition of (System) Architecture strongly resembles the Tarski definition of Model. It is also shown how the concepts of Architecture and Model can be logically separated by a small adjustment in the language of the current definition. Precise elementary examples are provided for conceptualisation of the approach offered. This position paper therefore offers a preliminary demonstration of how prose definitions can be complemented by a mathematical basis that can be used to improve the precision of the prose.
\end{abstract}
\begin{document}

\flushbottom
\maketitle
\thispagestyle{empty}

\section*{Introduction}

Although the concepts of Architecture and System have been linked together in mathematics, science and engineering for many decades, the precise definitions of these concepts and their relationships remain the subject of debate. For example, the diverse meanings of ‘architecture’ in Systems Engineering have been analysed using a soft systems approach \cite{WiKi+:10,EmBr+:12}. A similar approach was later used to analyse perspectives on the meaning of ‘system’\cite{SiDo+:17}. Despite the uncertain meanings, the application of Architecture and System concepts continues to grow apace across a wide range of disciplines. 

Questions of definition and meaning are not simply of academic interest because Architecture and System concepts are key to everyday practice and are foundational in international standardisation. As long ago as 2008 it was reported that there were more than 130 ISO standards with the word architecture in their title or abstract \cite{Be:08}. At that time, the Architecture Working Group (AWG) of the International Council on Systems Engineering (INCOSE) supported the standards community in analysing the plethora of standards relating to architecture - but without reaching any definitive conclusions about how they could be rationalised.

Within the standards community, the ISO/IEC Joint Technical Committee 1 (JTC 1), Sub-Committee 7 (SC 7), is responsible for software and systems engineering standards. Within SC 7, Working Group 42 is currently revisiting ISO/IEC/IEE 42010: 2011 on architecture description\cite{ISO42010}, and has already completed significant work on ISO/IEC/IEEE FDIS 42020\cite{ISO42020} on architecture processes and on ISO/IEC/IEEE FDIS 42030\cite{ISO42030} on architecture evaluation. The process and evaluation standards adopt a broader conception of architecture than previously, so that the standards apply to a range of objects including, but not limited to, traditional systems. As stated by Wilkinson\cite{Wi:16}, such a broadening of scope offers a unique opportunity to address the uncertainty of meanings and to bring together disparate disciplines within a single conceptual architectural framework.

Previous work highlights very strongly that a new approach is required to provide a stable basis for discipline and standards development in future. Following Wilkinson\cite{Wi:16}, we believe that this approach must:
\begin{itemize}  
  \item Focus on architecture as a primary concept, rather than as a feature of a system;
  \item Be universally applicable to all types of systems; 
  \item Ensure that the concepts of architecture have a rigorous foundation. 
\end{itemize} 

This paper offers early insights from research being undertaken by the authors to achieve the above, which we believe could inform the standardisation work of ISO/IEC WG42.

\section*{Background}

It is appropriate to begin with a short review of definitions from relevant international standards. 

From ISO/IEC/IEE 42010: 2011\cite{ISO42010}, we find that architecture is defined in terms of \textit{"...concepts or properties of a system in its environment embodied in its elements, relationships..."}
whereas in ISO/IEC/IEEE 15288:2008\cite{ISO15288}, it is defined in terms of \textit{"...organization of a system embodied in its components, their relationships to each other, and to the environment...}. Both standards declare that only fundamental elements are included in architecture. Both standards also include \textit{principles} relating to design and evolution of the architecture as part of architecture.

The more recent ISO/IEC/IEEE FDIS 42020\cite{ISO42020} and 42030\cite{ISO42030} standards have a similar definition to their sibling 42010\cite{ISO42010} but with broader application, preferring to replace the word “system” with “architecture entity”.

In the case of ISO/IEC/IEEE 42010:2011\cite{ISO42010}, the standard contains UML-style diagrams purporting to define the various concepts used within the standard. The fact that these diagrams are very much open to interpretation may explain why the uptake of the standard by practitioners is reported to have been limited.

\section*{A Different Approach to Conceptualisation}
All of the definitions in the key standards generally regard Architecture as a property of System (or some other entity). We regard architecture as a primary concept. Therefore, our approach begins with this concept. Breaking down the existing definitions, we see that there are three shared properties:
\begin{itemize}  
  \item Architecture should include principles of design and evolution;
  \item Architecture should include whatever is embodied in entities and the relationships between them;
  \item Architecture should be fundamental.
\end{itemize} 

Taking each of these in turn, we see that the first property is alluding to ‘engineered systems’ and therefore has no place in a general conception of architecture, which should apply to all types of systems. (We do not propose to broaden the concepts of Architecture to any exclusion of the concepts of System; rather we seek to explain how a general notion of architecture and a general notion of system can be related.) 

We suggest that the second property is concerned with the ‘structure’ of entities (i.e. elements or components) and their relationships; and to this we explicitly add the idea that ‘wholeness’ is emergent from this ‘structure’. In our conception of architecture, parts and the whole are equally important and both are required for a structure to be architectural. 

Finally, to say that architecture should be fundamental; we understand this to mean that the entities and relations included within the architecture are necessary and sufficient for there to be an emergent whole – there are no superfluous entities or relations, and none are missing.

So using natural language, architecture can be understood in terms of “the necessary and sufficient structure of entities and their relationships that give rise to an emergent whole”. We refer to this kind of structure as Architectural Structure. 

This line of reasoning leads to a consideration that a System is a realisation (or instantiation) of an Architectural Structure. This is consistent with the concept that the integration of elements in a specified way results in a system\cite{ISO15288}. In other words, the system must be compliant with a specified architecture. In this way we consider that any realisation of an Architectural Structure results in a System.

\section*{Definitions}

With the understanding reached above, we are now in a position to make natural language but precise definitions of key concepts that can be readily interpreted in mathematical terms. This will provide the rigorous underpinning we seek. 
\begin{addmargin}[4em]{4em}
\vspace{5pt}
\hspace{\parindent}1. \textit{\textbf{Structure} is the junction and separation of the objects of a collection that expresses how the junctures and separations provide explanation of a defining property of the collection.}

\vspace{1pt}
2. \textit{\textbf{Semantic Structure} is a structure that formalizes the intended meaning of concepts expressed through natural or domain language (knowledge of something).}

\vspace{1pt}
3. \textit{An \textbf{Architectural Structure} is a semantic structure of elements and relationships that is necessary and sufficient to express the properties of the elements as a whole.}

\vspace{1pt}

4. \textit{An image of a referent of a concept that is interpreted into an Architectural Structure is called an Architectural Interpretation.}

\vspace{1pt}

5. \textit{An Architectural Structure is therefore an abstract unification of Interpretations.}

\vspace{1pt}

6. \textit{If a referent of a Concept is a system, then it can be realised as an Architectural Interpretaion of the Concept.}

\vspace{5pt}
\end{addmargin}

The last point above offers a relation between our concepts and the concept of System. The first two definitions of structure are rooted in the ontology proposed by Sowa\cite{So:00}, and refined by Dickerson \& Mavris\cite{DiMa:16}.  The second definition is explained further by Ji and Dickerson in a recently adopted standard \cite{JiDi:18}. These definitions are currently being formalised by research being undertaken by the authors. Some of the mathematical basics are outlined in Annex A. We note that these key mathematical terms are used: object, collection, and property. When we use the term `object' from a mathematical viewpoint, it should not be confused with other uses of the term. We further note that using the term `object' from a mathematical viewpoint is not intended to replace current usage of the terms `element' or `entity' in the current ISO standards. 

\section*{Relation between System Architecture and Model}
A strong resemblance between System Architecture and Model can be seen when these two concepts are considered from a mathematical viewpoint. The relevant formalism is rooted in the Tarski theory of models in mathematical logic \cite{Ta:54,Ta:55,BeSl:69} as adapted to the practice of engineering \cite{MaGr+:13,DiMa:11}. Put succinctly, a \textit{\textbf{Model}} in the Tarski theory is defined to be a relational structure (in the sense of mathematical set theory) that is the image of an injective mapping of one or more sentences in the first order predicate calculus (a sentence being a fully quantified well-formed formula). In other words, the model ‘faithfully’ realises the sentences. See Ji and Dickerson\cite{JiDi:18} for further explanation.

Therefore, when \textit{\textbf{(System) Architecture}} is defined as ``...concepts or properties of a system in its environment embodied in its elements, relationships..." we find that ‘elements and relations’ bear a striking resemblance to the relational structure which is the image of an injective mapping in the Tarski theory. This is further reinforced when the interpretation of a concept (the injective mapping) into the relational structure is considered as ‘embodiment’.

This is a preliminary demonstration of how prose definitions can be complemented by a mathematical basis that can be used to improve the precision of the prose. The formal viewpoint on the definitions reveals that the definitions of system architecture and model might be in need of a stronger logical separation. This issue could be resolved in a straightforward way by a small adjustment in the language of the current definition:

\textit{\textbf{(System) Architecture}} is a fundamental concept that embodies (interprets) concepts and properties of a system that have been expressed through elements and their relationships.

We note that `system' is the higher level (mathematical) object in this definition; and `elements and their relationship's are the lower level (mathematical) objects. A definition like this not only conforms to Tarski Model Theory; it also expresses how architecture is explanation of the whole in terms of its parts. 

Although it has not been the purpose of this position paper to propose a new definition of Architecture, it is clear that the refinement and elaboration of terms in existing standards can be made more pprecise, and therefore more applicable to domain users, when concepts expressed in prose (natural language) are complemented by a mathematical basis for the concepts involved.

\section*{Examples}

Two examples are provided to illustrate the ideas explained above. Note that although these appear to be simple examples, they contain great richness and complexity.

\subsection*{Architecture of a System of Equations}
The first example is a simple mathematical one using linear simultaneous equations (see Annex A for a more extensive discussion of the mathematics). In this example, a collection of two elementary equations is defined: 
\begin{eqnarray}
\left\{
  \begin{array}{ll}\label{Eqn1}
  E_1: x_1+3x_2 = 10\\
  E_2: 2x_3-x_4 = 6\\
  \end{array}
\right.
\end{eqnarray}
These are then linked by specifying an interrelationship of the variables, for example: 
\begin{eqnarray}
\left\{
\begin{array}{ll}\label{Eqn2}
x_3 = x_1  \\
x_4 = x_2\\
\end{array}
\right.
\end{eqnarray}

Equation \ref{Eqn1} now represents a pair of linear simultaneous equations, such that they form a system of equations. This can be generalised into a 2-by-2 array using matrix and vector notation:
\begin{equation}\label{Eqn3}
\mathbf{A}\mathbf{x}=\mathbf{b}
\end{equation}
where
\[
\mathbf{A} = 
\begin{bmatrix}
	a_{11}       & a_{12}  \\
	a_{21}       & a_{22}
\end{bmatrix};\indent
\mathbf{x} = \begin{bmatrix}
	x_{1}        \\
	x_{2}     
\end{bmatrix};\indent
\mathbf{b} = \begin{bmatrix}
	b_{1}        \\
	b_{2}     
\end{bmatrix};
\]

Using our definitions, the (mathematical) objects that form the collection of equations are the variables and constants in Equation \ref{Eqn1}, their algebraic relationships (expressed through juxtaposition), and the relationships of equality. In a formal set-theoretic sense, the collection could be defined by the mathematical symbols. The interrelationship of the equations defined by the equalities in Equation \ref{Eqn2} is given as a method to form a system of equations from the collection of two otherwise unrelated equations in Equation \ref{Eqn1}. Thus, it is a defining property of the collection of equations that expresses when they have become a system of equations.

Equation \ref{Eqn3} represents a structure that separates the mathematical symbols into three groups: the variables ($\mathbf{x}$), the constants which are coefficients of the variables ($\mathbf{A}$), and those that are not ($\mathbf{b}$). [We note that each of these three groups has further structure.]  Two junctions join the three groups. The first is the juxtaposition of $\mathbf{A}$ and $\mathbf{x}$; and the second is the equality of $\mathbf{Ax}$ and $\mathbf{b}$. These junctions and separations express the defining property of the collection as a system of equations. Equation \ref{Eqn3} then clearly represents a structure as we have defined the term.

Furthermore, this is a semantic structure. It formalizes the intended meaning of the concept of a linear system of equations in terms of the domain language of mathematics.

Thus, only one thing remains to demonstrate that the matrix representation in Equation \ref{Eqn3} represents an Architectural Structure: specifically, that the semantic structure is necessary to express the defining property of the collection as a system of equations. We argue this logically in contrapositive. The defining property is expressed algebraically in Equation \ref{Eqn2}. If this fails, then either $x_3$ is distinct from $x_1$ or $x_4$ is distinct from $x_2$. In either case, the matrix multiplication not only fails; it is also undefined. Consequently, the semantic structure in Equation \ref{Eqn3} cannot express Equation \ref{Eqn1} if the defining condition is violated. We therefore conclude that Equation \ref{Eqn3} indeed represents an Architectural Structure.

Finally, Equation \ref{Eqn1} (under the condition of Equation \ref{Eqn2}) is clearly seen to be a realisation of the Architectural Structure by interpreting the elements of the matrix and vectors in Equation \ref{Eqn3} in the obvious way ($a_{11} = 1$, etc.). Thus, the (concrete) system of equations in Equation \ref{Eqn1} (when subjected to Equation \ref{Eqn2}) is indeed the result of realising an Architectural Structure. 

\subsection*{Architecture of a Torch}
A different example is provided by the concept of a Torch (see Annex B for further detail). We take the ‘essence’ of Torch to be a “hand-held source of illumination”. Several dictionary definitions match this concept:
\begin{enumerate}  
  \item A stick attached to a roll of hessian dipped in wax and set alight
  \item A large candle in a holder
  \item A handheld electric lamp, powered by batteries (UK)
\end{enumerate} 

These definitions are describing different types of physical Torch system. Some formalisation is required to transform each of these natural language definitions into more precise Architectural Interpretations. Even now, however, they appear conceptually similar and taken together they imply that there may be a more abstract concept of Torch within reach. This more abstract concept would be capable of unifying the formalised Architectural Interpretations. As discussed earlier, we term this  abstract unification of interpretations an Architectural Structure. 

By way of illustration, the Torch (1) definition can be developed for formalisation as follows:
\begin{itemize}  
  \item \textbf{Torch (1) description}: Torch (1) is a hand-held source of illumination made from a roll of hessian soaked in wax attached to a stick.
  \item \textbf{Stick}: A long thin piece of wood. In a Torch, a Stick is used to provide a rigid hand-hold away from the flame.
  \item \textbf{Wax impregnated hessian roll}: Combines both wax and hessian in a roll structure.
  \item \textbf{Hessian}: A piece of Hessian cloth. In a Torch, a Hessian Roll provides a wick for the molten wax.
  \item	\textbf{Wax}: A solid combustible material. In a Torch, Wax is the stored energy source.
  \item \textbf{Flame}: A chemical reaction that converts a combustible material to light, heat and by-products. In a Torch, the Flame converts Wax to light.
\end{itemize} 

This definition can be used to abstract fundamental concepts and their relations. For example, the definition Torch (1) is characterised physically, as shown in Figure \ref{fig1}:

\begin{figure}[h!]
\centering
\includegraphics[height=0.25\linewidth]{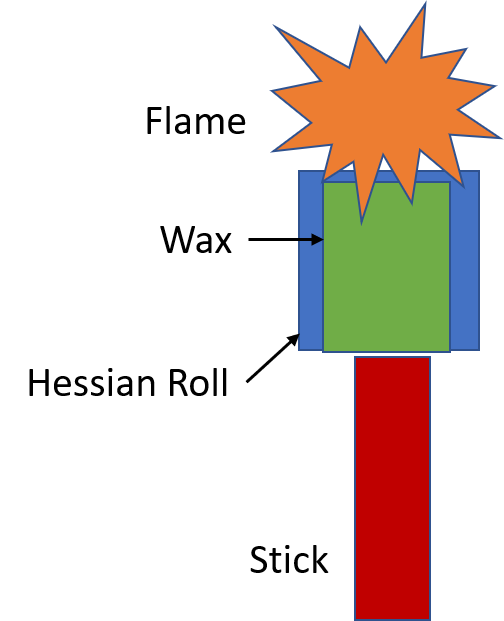}
\caption{Torch (1) – Fundamental Physical Concepts}
\label{fig1}
\end{figure}

\begin{figure}[h!]
\centering
\includegraphics[width=0.8\linewidth]{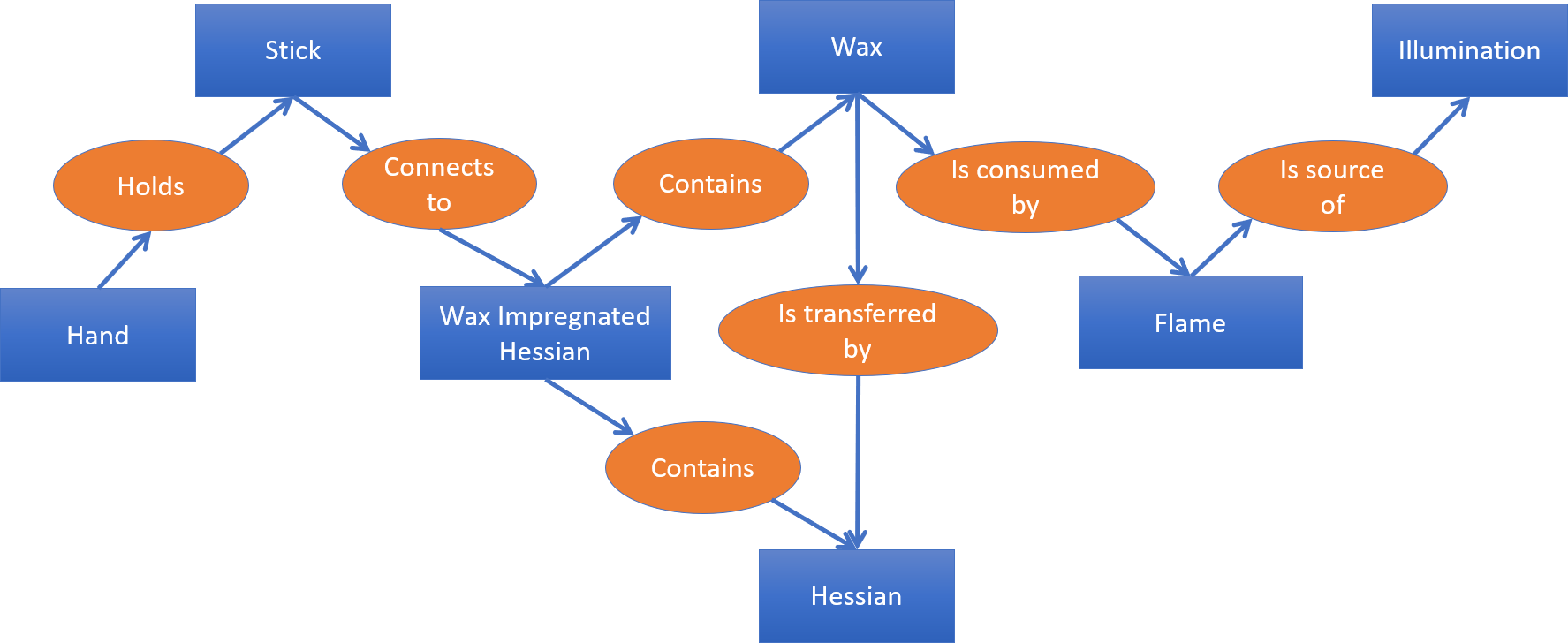}
\caption{Torch (1) – Architectural Interpretation: Fundamental Physical Concepts and their Relations
}
\label{fig2}
\end{figure}

We can now define a conceptual graph (see Annex B) for the Torch, as shown in Figure \ref{fig2}, containing the fundamental concepts of the Torch as described above, but also now identifying and formalising the relations between the concepts. This must be done minimally and holistically, as described earlier, to makes the conceptual graph an Architectural Interpretation for the physical realisation defined in Torch (1). As an Architectural Interpretation, here are no superfluous entities or relations in this conceptual graph, yet it manages to capture the ‘essence’ of “hand-held source of illumination”.

\begin{figure}[h!]
\centering
\includegraphics[width=0.8\linewidth]{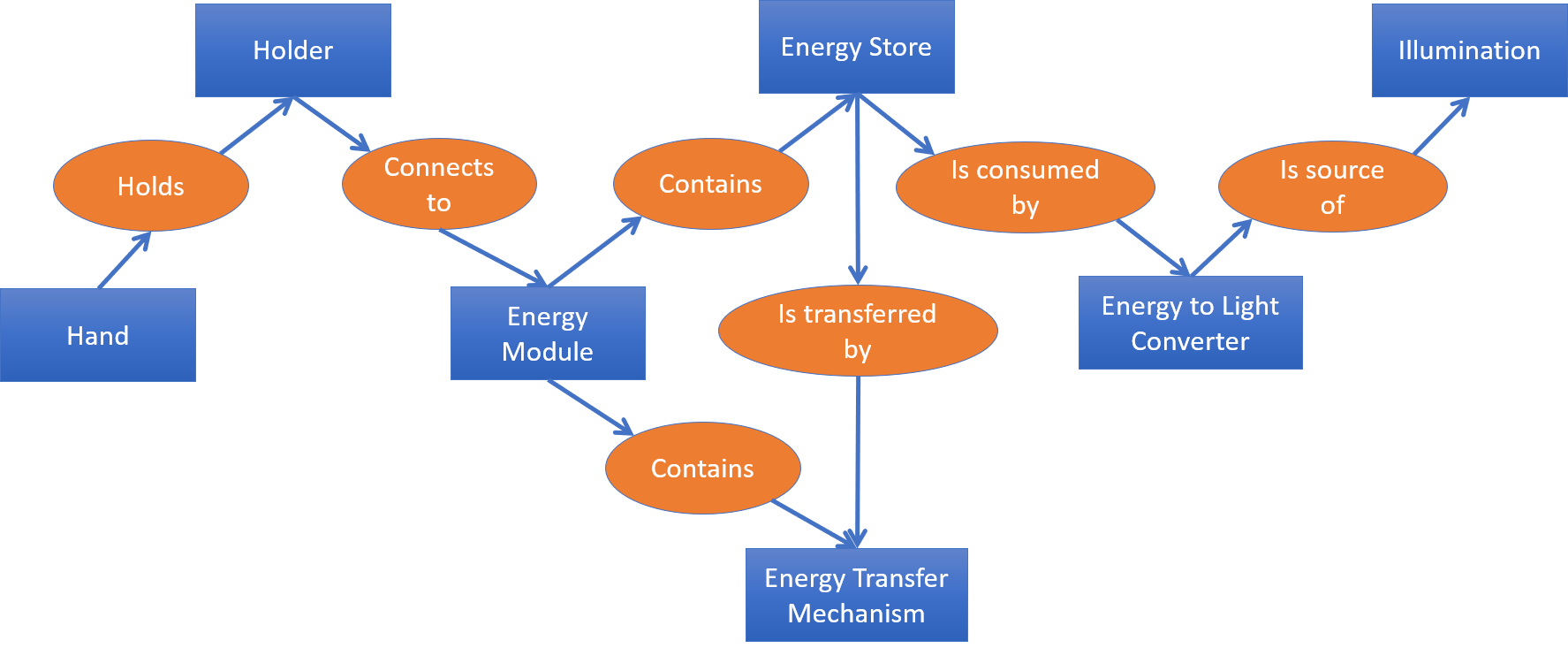}
\caption{Torch Architectural Structure:
Unification of Interpretations in an Architectural Structure
}
\label{fig3}
\end{figure}

A further process of abstraction, for example functional abstraction, can be followed to determine an Architectural Structure that covers several of the alternative Architectural Interpretations for Torch (see Annex B), as shown in Figure \ref{fig3}.

As an Architectural Structure, this is an abstraction that unifies multiple Torch Architectural Interpretations; it successfully captures the emergent abstract ‘torchiness’ we first identified – namely the hand-held portable source of illumination. The individual Architectural Interpretations can be realised as physical Torch systems built in accordance with the Concept.

Annex B develops this example further, explaining how the concepts are related to the mind’s semantic network and showing how the Architectural Structure might evolve in a practical engineering process.

\section*{Summary and Further Work}
It is our position that the definition of architecture should be complemented by the concepts of mathematics, science and engineering – with a suitably robust underpinning. We recommend that ISO WG 42 considers our proposals for the definitions of Architectural Structure and Architectural Interpretation with their relations to System as a way of providing the 420xx series of standards with a degree of formality hitherto lacking. Not only would this build on what has been achieved in previous standards, it would provide a pathway for further developments.

\bibliography{sample}

\section*{Annex A: Understanding the Concepts of System, Architecture, and Structure}
The purpose of this annex is to elaborate the mathematical basis of the simple abstract exemplar that was used as reference in the main text to discuss and reason about the concepts and terminology of System and Architecture, and models associated with the two terms. Of particular interest is the application of the exemplar to a demonstration that a meaningful concept of System aligned with the ISO/IEC/IEEE 15288: 2015\cite{ISO15288} can be refutable. 

The starting point is an Essential Systems Idea: to understand something of interest in context as a unified whole comprised of interrelated parts. Neither the whole object nor its parts are the sole subject of interest. 
When an object can be understood in this way, it will be referred to as a system.

\subsection*{Essential Mathematics}
The essential mathematics needed to reason logically about the essential idea above is the theory of classes and sets. In general, a \textit{class} $C(p)$ consists of (is the collection of) all objects having the property $p$. The objects of the class are also referred to as members of the class. The terms \textit{object}, \textit{collection}, and \textit{property} are undefined terms whose meaning are determined by the axioms of set theory, such as the Bernays-Gödel-von Neumann axiomatic set theory.  

Sets are a special type of class. Specifically, only those classes that can be members of a class are defined to be sets. This technical construct is used in order to avoid antinomies that can arise in class theory, such as the Russell paradox. Heuristically, a class can thought of as \textit{type} of object; and a \textit{set} can be thought of as a single collection of \textit{entities (objects that admit logical existence)}. Technically, a set can refer to physical objects but in itself, it is not a physical object. A set is an abstraction.

Sets are therefore normally discussed in the context of a predefined Universal Set, which is denoted as $U$. The process of defining a set ${S}$ is as follows: if $p$ is a property such that each element of $U$ either does or (exclusively) does not have the property $p$, then all of the elements $x$ belonging to $U$ that have the property $p$ form a set, which is denoted by $S = \left\{x\in U: p(x)\right\}$. The term \textit{property} is associated with a declarative statement, $p(x)$; as in the Propositional Calculus, e.g. ‘$x$ has the property $p$’. The objects which are members of a set are referred to as \textit{elements} of the set.

This level of understanding of set theory will be sufficient to to reason logically about the Essential Systems Idea.

\subsection*{Reasoning about the Concept of System}
It is clear from the discussion above that a system is a collection of objects. Furthermore, in order for the collection to have a logical existence it must be a set. This raises the question: what property $p$ must a collection of objects have in order to be defined as a system?

The Essential Systems Idea, as simple as it appears to be, is actually rich with semantics and relations between the terms that require further definition if something as precise as specifying ‘\textit{the defining property $p$ for a set to be a system}’ can be accomplished. It is important to note that this property actually might be a collection of properties.

The determination of such a defining property is a subject of Systems Thinking; and is beyond the scope of these notes. Therefore, the simple abstract example will be used to reason about the Concept of System.

\subsubsection*{Systems of Equations}
A defining property for System must be able to support the yes-no decision for when \textit{a collection of equations} can be defined as a \textit{system of equations}. Consider the two elementary equations provided in Equation \ref{Eqn1} in the main text:

The set, of course, is $\{E_1, E_2\}$, where $x_i$ are variables, and $a_{ij}$ and $b_i$ are constants. What property would make this set into an unified whole comprised of interrelated parts? As written, these two equations have no defined relationship.

\subsubsection*{A Defining Property}
One property of the equations that would establish a [system] relationship is that one or more of the variables are shared, e.g. $x_3 = x_1$ and $x_4 = x_2$. This particular case of shared variables results in a system of two linear equations in two variables. Any further discussion of the concepts and terminology of System and Architecture in this annex will be limited to this exemplar.

One of the membership properties that qualifies a set of equations to be a system of equations is expressed in terms of the interrelations between the elements of the set (in this case, a collection of equations); specifically the sharing of one or more variables.

\subsubsection*{Interrelationship}
This is significant because this particular (membership) property is not expressed in terms of the set definition process above, i.e. properties of the form $p(E)$ where $p$ is the defining property and $E$ is a candidate for membership. Rather, the property is of the form $p(x_i, x_j; E_1, E_2)$, i.e. based on interrelationship. 

\subsubsection*{Unified Whole}
Also, in the Essential Systems Idea, the term ‘unified’ somehow binds the parts and their relationships with the whole (as well as with each other). The property of equations sharing variables is a type of binding that should qualify as unification. This will be discussed further in ‘Architectural Structure’ below.

If the Essential Systems Idea is accepted as a starting point for reasoning about the Concept of System; then at this point, with the help of some elementary set theory, it has been possible to consider the terms in the \textit{first sentence} of the idea in a simple but precise way to understand what criteria might enable us to decide when a set of elements might be understandable as a system of elements.

The \textit{second sentence} of the idea has not really been considered yet: Neither the whole object nor its parts are the sole subject of interest
\subsection*{Reasoning about the Concept of Architecture}
Reasoning about the second sentence in the Essential Systems Idea is primarily a concern of Architecture. The concern is how the exemplar system (of equations) be understood as a whole without losing the details of the parts (i.e. details of the equations).

\subsubsection*{Matrix Representation}
Elementary algebra does this using matrix theory. It begins by arranging the coefficients into a 2-by-2 array, which is denoted $\mathbf{A} = [a_{ij}]$. This is referred to as a matrix. Similarly, matrices can be arranged for $\mathbf{x} = [x_i]^T$ and $\mathbf{b} = [b_i]^T$. The superscript $T$ indicates that the array has been transposed from a row vector into a column vector.

Equations \ref{Eqn1} can then be written in matrix algebra as shown in Equation \ref{Eqn3} in the main text when they are subjected to Equation \ref{Eqn2}

The system of equations can then be understood as a whole without disregarding the parts and their details; and the parts can understood without disregarding the whole. 

\subsubsection*{Architectural Structure}
The matrix representation of the system of equations, i.e. using Equation \ref{Eqn1} with the interrelationship in Equation \ref{Eqn2}, has separated the constant terms in the equations from the variable terms; and has joined these components back together through matrix multiplication and matrix equality. If the purpose of the system of equations is to find solutions for $x_1$ and $x_2$ that simultaneously satisfy both equations in Equation \ref{Eqn1}, then matrix algebra can be used to do this using Equation \ref{Eqn3}. 

\subsubsection*{Summary of Architecture}
Note that a system architecture has not been specified; but rather an Architectural Structure has been defined. The Architectural Structure expressed in Equation \ref{Eqn3} is an interpretation of the Concept of Linear Equation; and the System Model expressed in Equation \ref{Eqn1} with the interrelationship in Equation \ref{Eqn2} is seen as an instance of the Architectural Structure.

Architectural Structure is then a primary concept for system definition that allows something of interest (in this case, the system of equations) to be understood in a way where neither the whole object nor its parts are the sole subject of interest.

\subsection*{Summary of Systems Thinking on the Concept of System, Architecture, and Structure}
With a modest level of effort and formalism, it has been possible to validate the terms and concepts expressed in the Essential Systems Idea using an elementary exemplar. This exercise should make clear that it is helpful to start at an intuitive level of description. Even at the subsistent level of understanding of the Concept of System explored in the exemplar, there is a deep structure in the simple words of a concept such as a system that are expressed in the syntax and semantics of a surface structure such as natural language. However, the precision and mathematical basis of the simple example provide conceptualization of our approach that demonstrates its methods of application and its utility\cite{DiJi:16}.

\section*{Annex B: Using Conceptual Structures to understand Architectural Interpretations and Architectural Structure}

The purpose of this annex is to provide an intuitive illustration of the application of Architectural Structures and Architectural Interpretations to the Architecture-Centric Systems Engineering process. The aim of the example is to inform the ISO WG42 in its thinking about revisions to ISO/IEC 42010.

\subsection*{Conceptual Structures}
\begin{figure}[ht]
\centering
\includegraphics[height=0.3\linewidth]{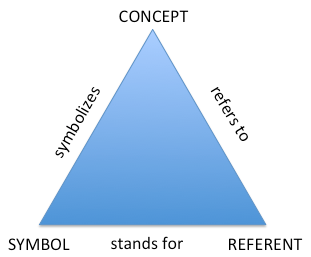}
\caption{The `meaning triangle'}
\label{FigB1}
\end{figure}

We draw on the framework of Conceptual Structures described by Sowa\cite{So:83}. This is essentially an approach to representing knowledge inside computer systems but it is more general and can be applied more widely. An abstract syntax and model-theoretic semantics for Conceptual Graphs are provided in ISO/IEC 24707:2018\cite{ISO24707}. Within the Sowa approach, Concepts are related to Symbols and to Referents in a ‘meaning triangle’, as depicted in \ref{FigB1} and described by Ogden and Richards\cite{OgCh+:23}.

Here, Symbol is a word representing an implicit Concept. The extension of the word is all of the things it applies to – which can be thought of as instances of the Concept. In other words, Concept is an \textit{intentional} abstraction of the common properties of the Referents to which is refers.

According to Sowa, abstract concepts acquire meaning not through direct associations with sense impressions (percepts) but with a vast network of relationships that ultimately links them to concrete concepts. The collection of all of the relationship that concepts have to other concepts, to percepts, to procedures and to motor mechanisms is called the \textit{semantic} network.

Thus, a conceptual graph has no meaning in isolation. It is only through the semantic network that its concepts and relations are linked to context, language, emotion and perception. In other words, what gives a conceptual graph meaning is its interpretation via a mind’s semantic network. As the semantic network is self-organising, it can adapt and change itself through external influences - and via nothing more than self-reflection.

Some of the semantic network can be formalised in a way that makes the internal workings of the network at least partly explicit. We can think of this ‘explication of the implicit’ as the manifestation of concepts and conceptual relations (as brought together in conceptual graphs). Making concepts and their relations explicit allows us to apply logical reasoning to them – which is the basis of mathematics, science and engineering.

The following shows the notation used by Sowa for conceptual graphs:
\begin{center}
[Concept 1] $\rightarrow$ (Conceptual Relation  $1\rightarrow2$)  $\rightarrow$ [Concept 2]
\end{center}
This is interpreted as meaning a concept, [Concept 1], is related via a conceptual relation (Conceptual Relation $1\rightarrow2$) to another concept, [Concept 2]. Each concept and conceptual relation is further defined in terms of their types and any constraints applying to them (for example a certain type of concept is always related to another type of concept by a particular conceptual relation). This is equivalent to having a limited knowledge domain captured in terms of concepts, conceptual relations and their definitions.

\subsection*{Elaboration of the Concept of ‘Torch’}
Torch has a variety of meanings cited in dictionaries, including several related to physical objects whose purpose is to create illumination, including:
\begin{enumerate}  
  \item A stick attached to a roll of hessian dipped in wax and set alight
  \item A large candle in a holder
  \item A handheld electric lamp, powered by batteries (UK)
\end{enumerate}

Some formalisation is required to transform each of these natural language definitions into more precise Architectural Interpretations. Even now, however, they appear conceptually similar and taken together they imply that there may be a more abstract concept of Torch within reach. This more abstract concept would be capable of unifying the formalised Architectural Interpretations. We term a concept that achieves the abstract unification of interpretations an Architectural Structure.

To see how this works, we analyse the concept of Torch in terms of the concepts relating to torches drawn from our knowledge base (or ontology) – which could be the dictionary mentioned above. We would expect to find the following terms defined as follows:

\textbf{Torch (1)}: Is made from a roll of hessian soaked in wax attached to a stick.
\begin{itemize}  
  \item {[Stick]}: A long thin piece of wood. In a Torch, a Stick is used to provide a rigid handhold away from the flame.
  \item {[Wax Impregnated Hessian Roll]}: Combines both wax and hessian in a roll structure. 
  \item {[Hessian]}: A piece of Hessian cloth. In a Torch, a Hessian roll provides a wick for the molten wax.
  \item {[Wax]}: A solid combustible material. In a Torch, Wax is the stored energy source.
  \item {[Flame]}: A chemical reaction that converts a combustible material to light, heat and by-products. In a Torch, the Flame converts Wax to light.
\end{itemize}

\textbf{Torch (2)}: Is made from a candle attached to a holder. In this case:
\begin{itemize}  
  \item {[Holder]}: A metal or wooden component with an attachment for the candle and shaped element suitable for hand-holding.
  \item {[Candle]}: A wax cylinder enclosing a wick. 
  \item {[Wick]}: A piece of string. In a Torch, the wick provides a means of transferring molten wax to the flame.
  \item {[Wax]}: A solid combustible material. In a Torch, Wax is the stored energy source.
  \item {[Flame]}: A chemical reaction that converts a combustible material to light, heat and by-products. In a Torch, the Flame converts Wax to light.
\end{itemize}

\textbf{Torch (3)}: Is made from a torch body containing batteries connected to a light bulb.
\begin{itemize}  
  \item {[Torch body]}: Acts as a handle as well as a convenient enclosure for the batteries.
  \item {[Batteries]}: Chemical devices for storing electrical energy.
  \item {[Connecting Wires]}: Connect the battery terminals to the electric light bulb.
  \item {[Light Bulb]}: Converts electrical energy to light energy, with by-products (e.g. heat).
\end{itemize}

All of the terms in these definitions would also be defined in the dictionary. The dictionary is therefore a self-referential closed system of definitions. The definitions are given meaning only when interpreted by a mind.

\subsection*{Reverse Architecting of Architectural Interpretations}
From the dictionary definitions we can develop an abstracted set of concepts and conceptual relations that correspond to the knowledge base.

A first level of abstraction identifies the fundamental physical concepts and relates them to each other. This has been depicted in Figure \ref{fig1} in the main text. 

This allows us to define a Torch conceptual graph containing the fundamental concepts and relations of the Torch described, which is therefore an Architectural Interpretation, as depicted in in Figure \ref{fig2} in the main text.

Similarly, for Torch (2), the fundamental physical concept is shown in Figure \ref{FigB2} below.

\begin{figure}[h!]
\centering
\includegraphics[height=0.3\linewidth]{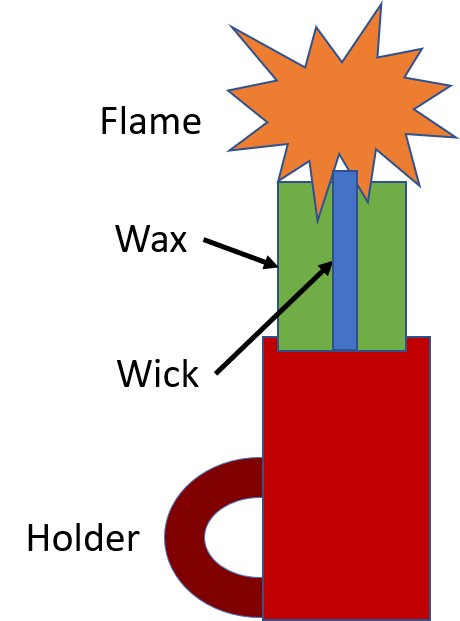}
\caption{Torch (2) – Fundamental Physical Concepts}
\label{FigB2}
\end{figure}

\begin{figure}[h!]
\centering
\includegraphics[width=0.8\linewidth]{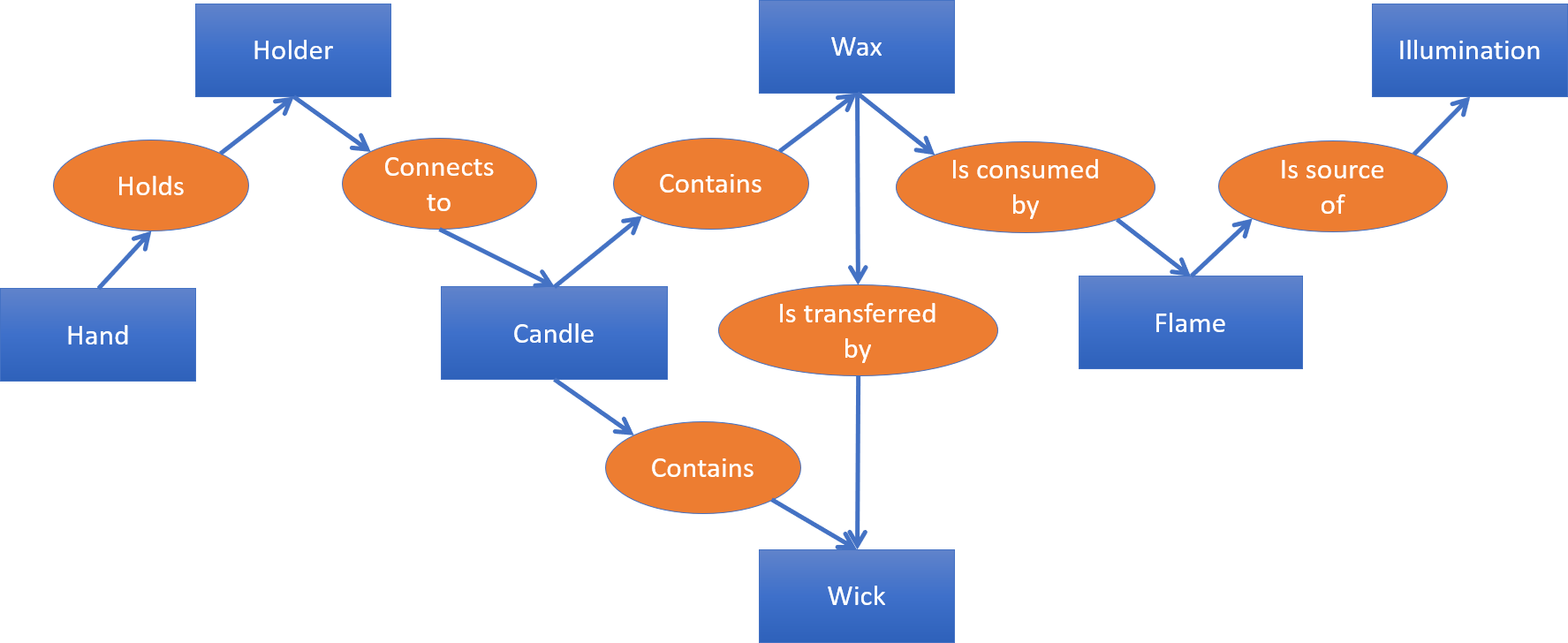}
\caption{Torch (2) – Architectural Interpretation: Fundamental Physical Concepts and their Relations
}
\label{FigB3}
\end{figure}

As before, the fundamental physical concepts can be related to each other to generate an Architectural Interpretation, which is depicted in Figure \ref{FigB3}

\subsection*{Unification of interpretations via an Architectural Structure}
\begin{figure}[h!]
\centering
\includegraphics[height=0.3\linewidth]{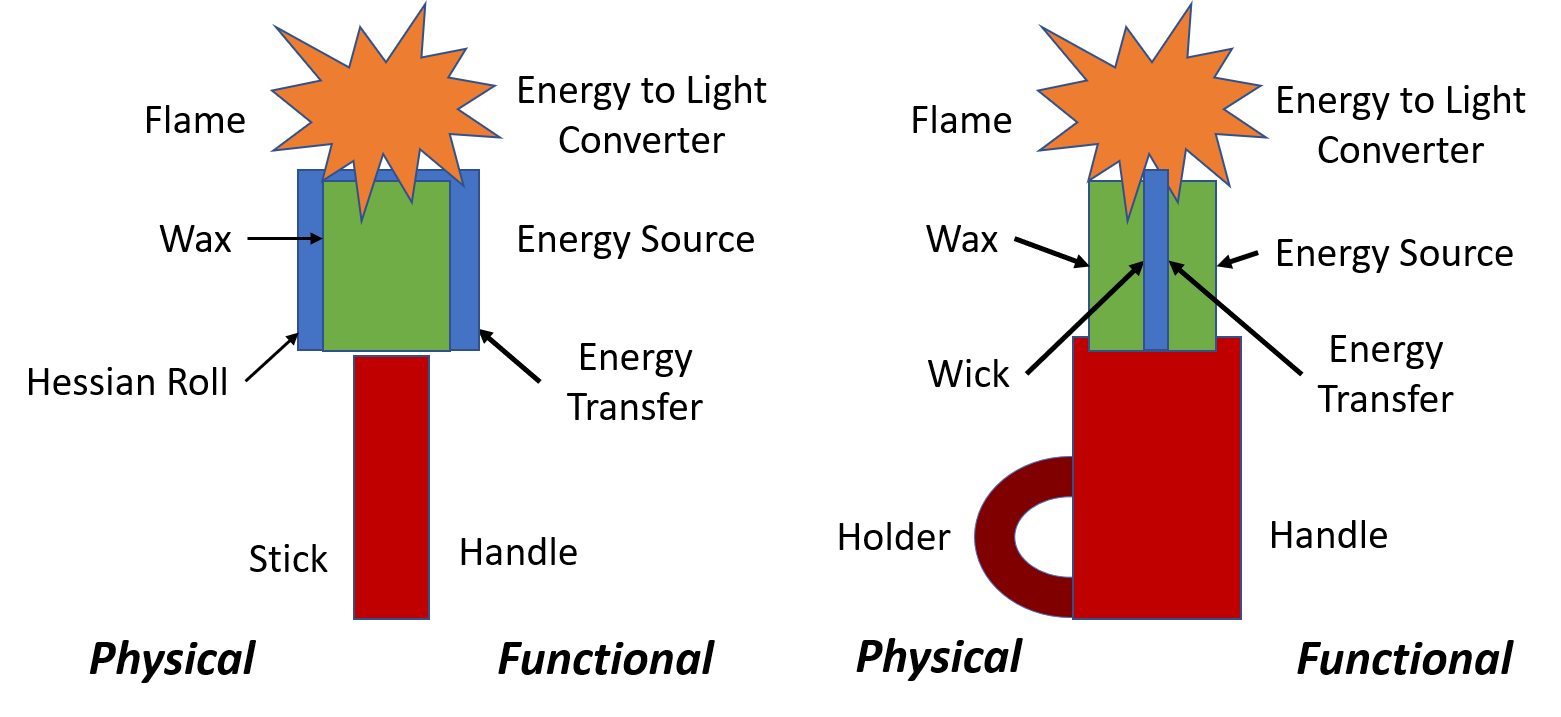}
\caption{Torch (1) (left) and Torch (2) (right) – Fundamental Functional Concepts}
\label{FigB4}
\end{figure}

A second level of abstraction is possible – in this example a functional abstraction based on the purpose of each physical element as part of a Torch (See Figure \ref{FigB4}). 

This leads to a unification of our first two Architectural Interpretations for Torch to yield an Architectural Structure for Torch depicted in Figure \ref{fig3} in the main text.

As an Architectural Structure, this is an abstraction that unifies multiple Torch Architectural Interpretations; it successfully captures the emergent abstract ‘torchiness’ we first identified – namely the hand-held portable source of illumination. The individual Architectural Interpretations can be realised as physical Torch Systems built in accordance with the appropriate Concept.

\subsection*{Architectural evolution}
\begin{figure}[h!]
\centering
\includegraphics[height=0.25\linewidth]{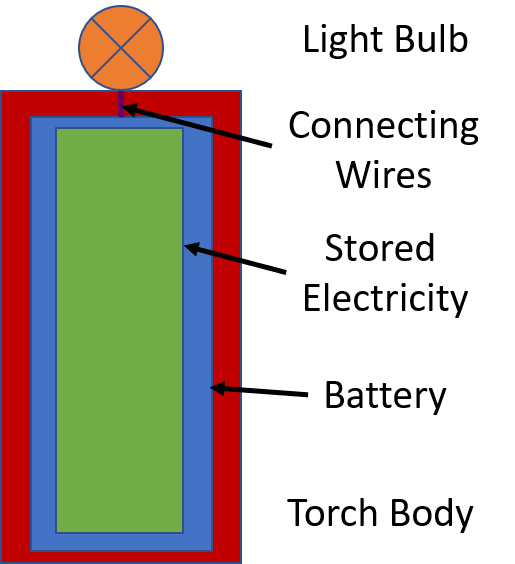}
\caption{Torch (3) – Fundamental Physical Concepts}
\label{FigB6}
\end{figure}

Working now with the Torch definition for the hand-held electric torch (which in American English might be recognised as a flashlight), we can define a Torch physical concept using the schematic presented in Figure \ref{FigB6}.

\begin{figure}[h!]
\centering
\includegraphics[width=0.8\linewidth]{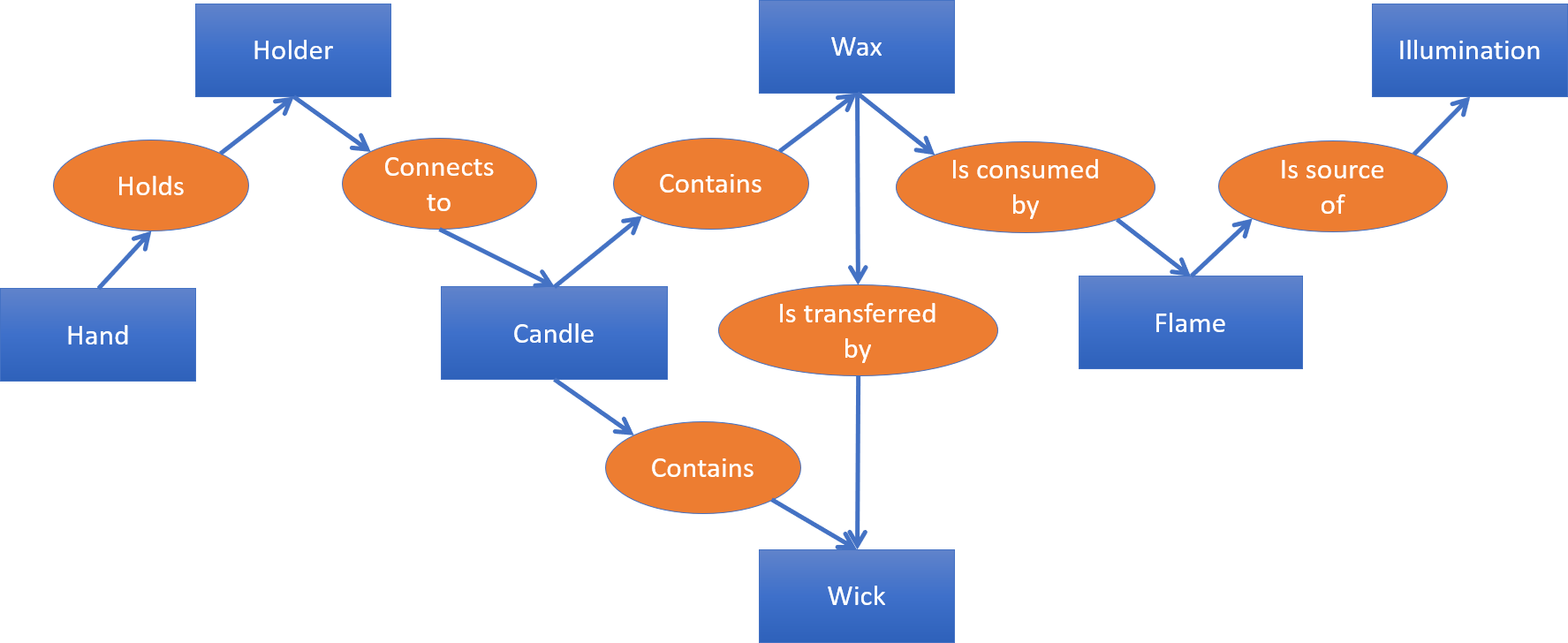}
\caption{Torch (3) – Architectural Interpretation:
Fundamental Physical Concepts and their Relations
}
\label{FigB7}
\end{figure}

As previously, we can define a conceptual graph, as depicted in Figure \ref{FigB7}, containing the fundamental concepts and relations of this Torch, so the graph is again an Architectural Interpretation. 
Note that in our first two Torch Architectural Interpretations there was a complete correspondence between the conceptual relations and equivalent concepts. For Torch (3), this is not the case, as the relation between the Torch Body and Battery (i.e. Encloses) is different to that between Stick and Wax-impregnated Hessian Roll and between Holder and Candle (i.e. Connects to). Similarly, the relation between Battery and Connecting Wires (i.e. Connects to) is different to that between Wax Impregnated Hessian and Candle and Wick (i.e. Contains). 

\begin{figure}[h!]
\centering
\includegraphics[width=0.8\linewidth]{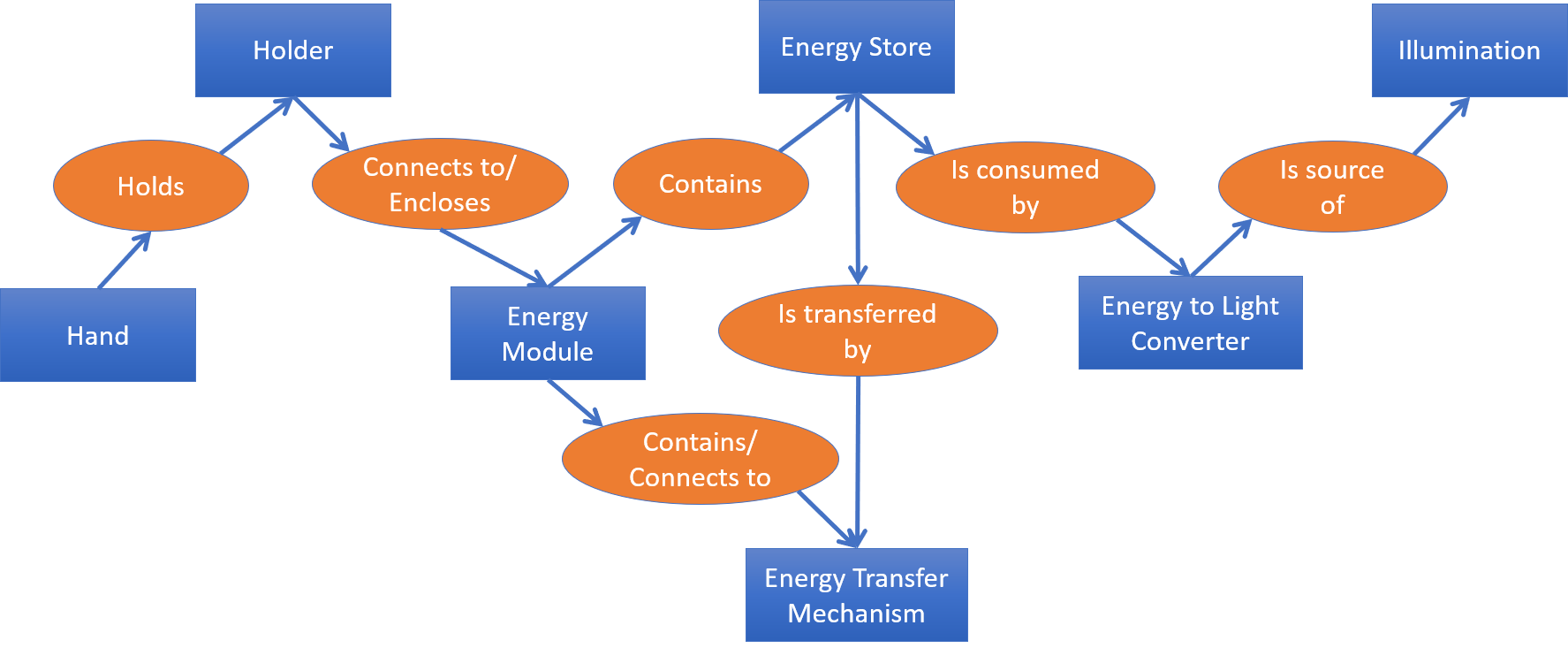}
\caption{Torch - Architectural Structure (version 2): Unification of Torch Interpretations in an Architectural Structure
}
\label{FigB8}
\end{figure}

This means that the Architectural Structure we had used previously needs to be generalised to ensure it can unify all of the Torch Architectural Interpretations. In this example, this is easily achieved by allowing some options for the relations in the Architectural Structure, as shown in Figure \ref{FigB8}.

Although this is a slightly artificial example chosen to illustrate a point, the general principle is that Architectural Structures are not fixed but tailored for a purpose. Any relevant discovery in mathematics, science or engineering is likely to have an impact on Architectural Structures. This could be manifested as changes to the fundamental concepts, conceptual relations or to the underlying graph. 

\subsection*{Summary of the use of Conceptual Structures in reversing architecting}

The summary of what we have done in this example can be described as follows:
\begin{itemize}  
  \item We took a Concept from a knowledge domain and analysed it to determine an Architectural Interpretation.
  \item We repeated this for a similar concept from a similar knowledge domain.
  \item We then took the two similar Architectural Interpretations and abstracted from them an Architectural Structure to unify them both.
  \item We took a similar third Concept from a similar knowledge domain and analysed it to determine a third Architectural Interpretation.
  \item We then in interpreted this Architectural Interpretation in terms of the previously unified Architecture Structure to determine discords between them. This allowed us to re-define the Architectural Structure to ensure it would apply to all defined Interpretations of [Torch].
\end{itemize} 

\end{document}